\documentclass[aip,floats,superscriptaddress]{revtex4}

\usepackage{graphicx}
\usepackage{epstopdf}
\usepackage{amssymb}
\usepackage{amsmath,bm}
\usepackage{psfrag}
\usepackage{epsfig}
\usepackage{float}
\usepackage{bm}
\newcommand{\angstrom}{\mbox{\normalfont\AA}}

\begin{document}

\title{Directed long-range transport of a nearly pure component atom clusters by the electromigration of 
a binary 
surface alloy
}

\author{Mikhail Khenner\footnote{Corresponding
author. E-mail: mikhail.khenner@wku.edu.}}
\affiliation{Department of Mathematics, Western Kentucky University, Bowling Green, KY 42101, USA}
\affiliation{Applied Physics Institute, Western Kentucky University, Bowling Green, KY 42101, USA}

\begin{abstract}
\noindent

Assuming a vacancy-mediated diffusion, a continuum model for electromigration-driven transport of an embedded atom cluster across a surface terrace 
of a phase-separating A$_x$B$_{1-x}$ surface alloy, such as fcc AgPt(111), is presented. Computations show that the electron wind carries 
the cluster over hundreds of lattice spacings and in the set direction, while the cluster grows and its purity improves during the drift. 
Impacts of the current density, the diffusion anisotropy, the magnitude and sign of the ratio of the effective charges $q_A/q_B$, and the jump frequencies 
ratio $\Gamma_A/\Gamma_B$ on the cluster's drift speed, drift direction, purity and shape are demonstrated.  


\end{abstract}

\date{\today}
\maketitle


\section{Introduction}
\label{Intro}

STM studies of the early 2000's \cite{GSASF,GSBH,AAFS} demonstrated that atoms in close-packed crystal surfaces of pure metals or alloys are highly mobile 
even at room temperature. For instance, Pd or In atoms incorporated into the topmost layer of Cu(001) substrate 
may frequently jump over distances larger than one lattice spacing, some as far as five lattice spacings. These jumps are enabled by a 
fast two-dimensional random  walk of vacancies. Of course, the observed two-dimensional motion of impurity atoms
reflects the diffusion of all copper atoms in the surface layer. Such vacancy-mediated diffusion 
of the surface atoms was compared to a giant atomic slide-puzzle \cite{GSSF}.    

Concerning this phenomenon, the general question one may be asking is this: 
Is it possible to channel, or bias, the random vacancy-mediated diffusion of individual surface atoms into a collective, 
directed transport across the surface by the application of the external field or force ?
In this communication, we answer positively. Using a continuum model based only on the classical diffusion and electromigration phenomenology 
\cite{Huntington,HoKwok,Mehrer}, we show that the ``electron wind" may transport, over distances of hundreds of lattice spacings and in the 
set direction on the surface, the compact clusters comprising of, say, predominantly B atom species. As the cluser is transported along the terrace, 
the persistant phase separation ensures the continuing in-flow of B atoms and the corresponding out-flow of A atoms from the cluster, 
until a 100\% purity is achieved. The basic necessary conditions are that a phase-separating instablity of the surface alloy is activated, 
and that prior to the onset of the electromigration, there exists a small seed cluster with a spiked concentration of B atoms. 
Such seed cluster may be artificially created by a local implantation of B atoms into A$_x$B$_{1-x}$ surface alloy, or
it may occur naturally in the said alloy.    

The model and the computations are presented in sections \ref{Model} and  \ref{CompositionEvolve}, respectively. 
The model extends the analysis in Ref. \cite{MySurfSci} and provides a self-consistent formulation, 
based on irreversible thermodynamics, of the vacancy-mediated embedded atom surface diffusion and electromigration. It is appropriate to remark here
that several useful models of surface electromigration-driven nanoisland motion and instabilities were published \cite{CMECPL,KKHV,DasM,DKM,KDDM,SV,THP}, 
however, these studies do not include alloys and the diffusing and migrating species in this case are the adatoms - 
thus the diffusion is by the direct exchange of lattice positions.

It is important to point out that the nature of a phase-separating instability is not highly important for the cluster transport on a surface. 
Any physical mechanism that enhances, in the non-explosive fashion, the initial localized phase inhomogeneity would suffice in the presented model, although
the drift speed of the cluster and the (time-dependent) cluster purity will be affected by the detailed workings of the instability.    
In this communication it is assumed that the surface alloy is thermodynamically unstable, that is, the instability is spinodal \cite{CH,C}. This is the simplest
phase-separating instability that can be manifested.

Literature on surface alloys is vast, however, the majority of studied alloys are thermodynamically stable. We came across the discussion of one fcc surface 
alloy system, AgPt(111) which is thermodynamically unstable \cite{RSBK}. 
These authors describe AgPt(111) surface alloy as
``... a solid solution with a positive enthalpy of mixing, which leads to phase separation". 
They find that the growth of Ag on Pt(111) 
results in compact Ag clusters of the average size 10$\angstrom$ embedded in the topmost Pt layer. (The clusters were not observed for 
a period of time at constant temperature, thus it is not clear whether they grow or not.) Such incomplete mixing within the surface layer is 
consistent with the phase diagram for a bulk AgPt alloy, which features a wide miscibility gap up to $T\sim1400$K \cite{SCP}. The clusters dissolve
into the topmost Pt layer only when the film is annealed at $T>650$K for about one minute. It is therefore possible that application of the 
electromigration to AgPt(111) surface alloy after Ag clusters were formed would result in these clusters drifting within the Pt surface layer, 
while simultaneously growing by phase separation.
In the paper we provide theoretical and computational support for precisely this scenario. It should be understood that despite us using specific published 
material parameters for AgPt surface alloy for computations and estimates, our model is applicable to other surface alloys that display 
qualitatively similar thermodynamics and kinetics. For example, Ni-based fcc alloys, such as NiCo and NiCr 
also have a positive enthalpy of mixing in the bulk phase \cite{ZSZ}, thus they are the potential candidates for displaying phase separation when realized 
as surface alloys.
 
We hope that the proposed transport effect of the electric current on the surface atom clusters could be demonstrated in a real-life experiment and 
then harnessed to enable the localized chemical functionality on metal surfaces, for instance, allow the creation of the mobile catalytic "hot-spot".

\section{The Model}
\label{Model}

\begin{figure}[H]
\vspace{-0.2cm}
\centering
\includegraphics[width=4.0in]{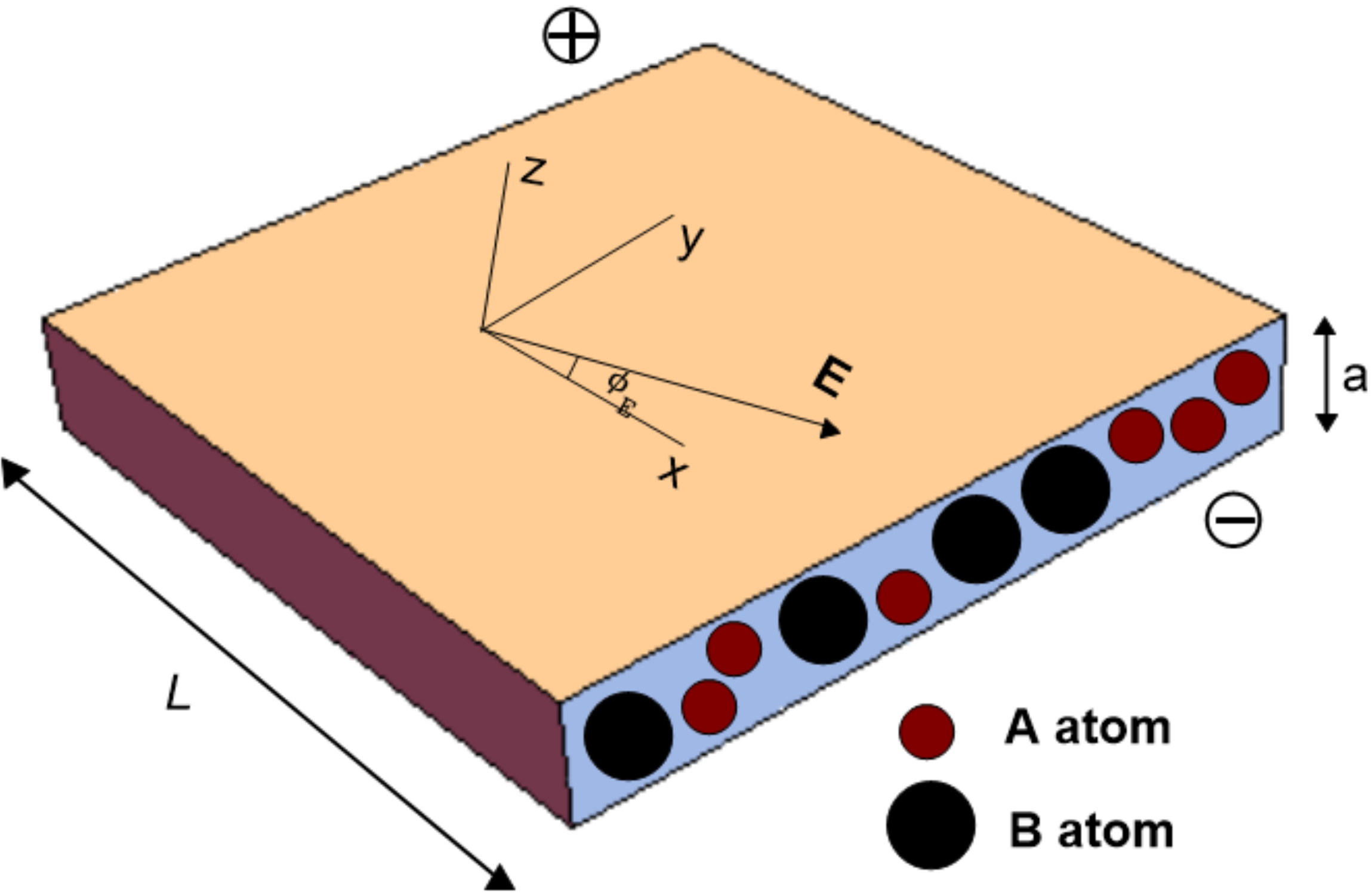}
\vspace{-0.15cm}
\caption{Schematic representation of a single layer surface binary alloy subjected to an external electric field $\bm{E}$. 
}
\label{Fig1}
\end{figure}

We consider a thin conductive film of a uniform composition A, whereby the impurity B atoms are mixed into the topmost surface layer, thus creating
a binary surface alloy. The film's surface morphology is comprised of the atomically flat terraces of width $L$ separated by steps. 
The electric potential difference (voltage) $V$ is applied to the opposite edges of the terrace, which results in
the electric field $\bm{E}=\left(E_0\cos{\phi_E},E_0\sin{\phi_E}\right),\; E_0=V/L$. 
Here $\phi_E$ is the angle that the electric field vector makes with the $x$-axis of the Cartesian reference frame, see Fig. \ref{Fig1}.
Note that the $x$ and $y$ axes are chosen along the principal diffusion axes.

Let $X_A$ and $X_B$ be the local time-dependent surface atom concentrations $(\left[X_A,\ X_B\right]=cm^{-2})$, and let $\nu_A$ and $\nu_B$ be the 
corresponding average (and constant) concentrations, that is, the number of A and B atoms per unit area in the as-prepared surface alloy prior to 
the onset of a phase separation and electromigration. Then
\begin{equation}
\frac{X_A}{\nu_A}+\frac{X_B}{\nu_B}=C_A+C_B=1, 
\label{sumC}
\end{equation}
where $C_A$ and $C_B$ are the dimensionless local time-dependent concentrations, or the local composition fractions.

%


The diffusion equations read  
\begin{equation}
\frac{\partial X_A}{\partial t} =   - \bm{\nabla} \cdot \bm{J}_A,\quad 
\frac{\partial X_B}{\partial t} =   - \bm{\nabla} \cdot \bm{J}_B, \label{C-eq}
\end{equation}
where 
$\bm{J}_A,\ \bm{J}_B$ are the components' fluxes measured in the laboratory reference frame, 
and $\bm{\nabla}=(\partial_x, \partial_y)$. We are modeling the alloyed terrace as the two-dimensional (2D) diffusion couple, where diffusion 
is mediated by vacancies (see the remarks in Sec. \ref{Disc} on comparing various diffusion rates). 
Thus to ensure the mass conservation,  according
to the classical framework \cite{Darken}
one introduces the intrinsic (lattice) 
fluxes of the components, $\bm{j}_A$ and $\bm{j}_B$, that are related to the fluxes in the laboratory frame via
\begin{equation}
\bm{j}_A = \bm{J}_A - \bm{v} X_A,\quad \bm{j}_B = \bm{J}_B - \bm{v} X_B, \label{Jlattice-eq}
\end{equation}
where $\bm{v}(x,y,t)$ is the velocity field \cite{Mehrer,Gusak}. The velocity field is measured in the laboratory frame and characterizes 
the velocity of the 2D surface lattice relative to the bulk, non-diffusing parts of the sample away from the interdiffusion area.

The conservation laws (\ref{C-eq}) then have the form:
\begin{equation}
\frac{\partial X_A}{\partial t} =   - \bm{\nabla} \cdot \bm{j}_A - \bm{\nabla} \cdot \left(\bm{v} X_A\right),\quad 
\frac{\partial X_B}{\partial t} =   - \bm{\nabla} \cdot \bm{j}_B - \bm{\nabla} \cdot \left(\bm{v} X_B\right). \label{C-eq1}
\end{equation}

Multiplying the first conservation law 
in Eq. (\ref{C-eq1}) by $1/\nu_A$, the second conservation law by $1/\nu_B$, adding, and using Eq. (\ref{sumC}) gives
\begin{equation}
\bm{\nabla} \cdot \bm{v} = -\bm{\nabla}\cdot (\frac{\bm{j}_A}{\nu_A} + \frac{\bm{j}_B}{\nu_B}). \label{divV}
\end{equation}
The trivial solution of this equation for $\bm{v}$ is assumed in the foregoing formulation:
\begin{equation}
\bm{v} = - \frac{\bm{j}_A}{\nu_A} - \frac{\bm{j}_B}{\nu_B}. \label{V}
\end{equation}
Eq. (\ref{V}) states that the 
lattice moves
in the direction opposite to the direction of the vector sum of the lattice 
fluxes $\bm{j}_A$ and $\bm{j}_B$. It is expected that in practice the deviations from this rule are very small and thus in the first approximation they 
can be neglected (\cite{Mehrer}, p. 198). If the deviations cannot be neglected, the analysis becomes considerably more complicated 
\cite{BGCM}. 
Typically, the vectors $\bm{j}_A/\nu_A$ and $\bm{j}_B/\nu_B$ (which are analogous to $V_A \bm{j}_A$ and $V_B \bm{j}_B$ for the bulk alloy, 
where $V_A$ and $V_B$ are the partial molar volumes \cite{BGCM}) point in the opposite directions and their magnitudes are nearly equal, thus the lattice velocity is 
very small. Yet, for bulk three-dimensional alloys it can be measured in the interdiffusion/Kirkendall experiments. In this communication, the focus is not on 
modeling the probable Kirkendall effect in the binary surface alloy, but on demonstrating the surface atom transport effect of the applied electric current. Thus we 
are not concerned with the computation of Eq. (\ref{V}), and it is presented only as part of the model formulation.

With $\bm{\nabla}\cdot \bm{v}$ given by Eq. (\ref{divV}) and $\bm{v}$ given by Eq. (\ref{V}), it is seen that the expanded conservation laws 
\begin{equation}
\frac{\partial X_A}{\partial t} =   -\bm{\nabla} \cdot \bm{j}_A - X_A\bm{\nabla} \cdot \bm{v} - \bm{v} \cdot \bm{\nabla} X_A,\quad 
\frac{\partial X_B}{\partial t} =   -\bm{\nabla} \cdot \bm{j}_B - X_B\bm{\nabla} \cdot \bm{v} - \bm{v} \cdot \bm{\nabla} X_B  \label{C-eq2}
\end{equation}
are now explicitly formulated in terms of the lattice fluxes. 
We now proceed to the formulation of these fluxes.


According to the classical, irreversible thermodynamics-based phenomenological model of electromigration in metal alloys \cite{Huntington,HoKwok}, 
\begin{eqnarray}
\bm{j_A} &=& -L_{AA} \left(\bm{\nabla} \left[\mu_A-\mu_V\right] -q_A \bm{E}\right) - L_{AB} \left(\bm{\nabla} \left[\mu_B-\mu_V\right] -q_B \bm{E}\right),\nonumber \\
\bm{j_B} &=& -L_{AB} \left(\bm{\nabla} \left[\mu_A-\mu_V\right] -q_A \bm{E}\right) - L_{BB} \left(\bm{\nabla} \left[\mu_B-\mu_V\right] -q_B \bm{E}\right).\label{j-eq2}
\end{eqnarray}
Here $L_{AA},\ L_{AB},\ L_{BB}$ are the kinetic transport coefficients, $\mu_A,\ \mu_B$ the chemical potentials of the components, 
$\mu_V$ the chemical potential of vacancies, and $q_A,\ q_B$ the effective 
surface charges of the atom species A and B, respectively. The effective charges include the electrostatic and the electron-wind contributions, and are defined as
\begin{equation}
q_A=\left(Z_A^{(e)}+Z_A^{(w)}\right)e,\quad q_B=\left(Z_B^{(e)}+Z_B^{(w)}\right)e,
\end{equation}
where $e<0$ is the electron charge. Note that, since the sums $Z_i^{(e)}+Z_i^{(w)}$ of the effective ion valences may be positive or negative, $q_A$ and $q_B$ also may be positive or negative.
When the magnitude of the electron-wind force is much larger than the magnitude of the electrostatic force, which is typical in surface electromigration 
phenomena \cite{BGWZ,REW}, 
the positive effective charge moves in the $-\bm{E}$ direction (equivalently, in the direction of the electron flow), that is, from the negatively charged anode to the positively charged cathode.
Also note that due to formulating the diffusion equations in terms of the lattice fluxes, the kinetic transport coefficients are uniquely determined and their 
number is reduced from four to three, in other words, the off-diagonal coefficients are equal, $L_{BA}=L_{AB}$ \cite{Huntington,HoKwok}. 
We further assume that the vacancy 
sources and sinks (such as the step edges) are sufficiently effective to keep $\mu_V$ constant, and also that the binding energies V-A and V-B are equal.
\footnote{This is supported by ab initio computation for some concentrated fcc alloys. For example, in Ref. \cite{ZSZ} it is shown that the formation 
energies of Ni and Co vacancies in Ni$_{0.5}$Co$_{0.5}$ alloy are nearly equal, and they exhibit the same trend as a function of the number of vacancy 
nearest neighbor pairs.} Together, these assumptions
allow to drop $\bm{\nabla} \mu_V$ from Eqs. (\ref{j-eq2}) \cite{Huntington}.

The chemical potentials in Eq. (\ref{j-eq2}) are due to alloy thermodynamics \cite{ZVD}:
\begin{equation}
\mu_A = \frac{C_B}{\nu_B}\left(\frac{\partial \gamma}{\partial C_A}-\frac{\partial \gamma}{\partial C_B}\right)-\frac{\epsilon a}{\nu_A} \bm{\nabla}^2 C_A,\quad
\mu_B = \frac{-C_A}{\nu_A}\left(\frac{\partial \gamma}{\partial C_A}-\frac{\partial \gamma}{\partial C_B}\right)-\frac{\epsilon a}{\nu_B}\bm{\nabla}^2 C_B, 
\label{base_eq4}
\end{equation}
where $\gamma$ is the free energy, 
$a$ the lattice spacing, and $\epsilon$ the Cahn-Hilliard gradient energy coefficient. Contributions to the chemical potentials from the compositional 
strain (e.g., strain that emerges because of the size mismatch of A and B atoms) are expected to be minor for AgPt surface alloy, since the size mismatch
is not significant - the radius of Ag atom is $1.65\times 10^{-8}$ cm, whereas  the radius of Pt atom is $1.77\times 10^{-8}$ cm \cite{CRR}. See References 
\cite{MySurfSci,MyMSMSE} for account of strain in a related model based on a generic diffusion setup.\footnote{By a generic diffusion setup we mean a setup
like Fick's diffusion equation, e.g. one where the details of diffusion are not considered.}

The free energy is chosen as \cite{RSN,LuKim}:
\begin{equation}
\gamma=\gamma_A C_A+\gamma_B C_B +k T \nu \left(C_A\ln C_A + C_B\ln C_B + H C_A C_B\right),
\label{gamma}
\end{equation}
where it is assumed $\nu_A=\nu_B=\nu$, $\gamma_A$ $(\gamma_B)$ is the surface energy of atomically thin surface layer that is composed of A (B) atoms, 
$kT\nu$ the alloy entropy, and the dimensionless number 
$H=\alpha_{int}/k T\nu$ measures the bond strength relative to the thermal energy $k T$. Here $\alpha_{int}$ is the enthalpy of mixing.
The terms in the parenthesis are the regular solution model. Note that when B atoms are deposited on a surface composed of A atoms, 
if B atoms have lower surface energy ($\gamma_B < \gamma_A$), then they tend to be confined to the surface layer; if $\gamma_B > \gamma_A$, 
then B atoms tend to be dissolved in the A-bulk \cite{RSN}.

Apart from the constant matrix $\bm{W}$, which will be introduced shortly, the kinetic transport coefficients have the forms derived from  
Manning's model of diffusion in nondilute alloys \cite{Manning1,RWM}:
\begin{eqnarray}
L_{AA}&=&\Gamma_B \Gamma C_A \frac{\lambda a^2\nu}{k T}\left(1-\frac{2\Gamma C_B}{\psi}\right)\bm{W},\quad 
L_{AB}=\Gamma_B \Gamma C_A C_B\frac{2\lambda a^2\nu}{k T\psi}\bm{W}, \nonumber \\ 
L_{BB}&=&\Gamma_B C_B \frac{\lambda a^2\nu}{k T}\left(1-\frac{2 C_A}{\psi}\right)\bm{W} \label{LAALABLBB},
\end{eqnarray}
where the dimensionless $\psi$ is given by
\begin{equation}
\psi = \frac{1}{2}\left(M_0+2\right)\left(\Gamma C_A+C_B\right)-\Gamma-1+2\left(C_A+\Gamma C_B\right)+
\sqrt{\left[\frac{1}{2}\left(M_0+2\right)\left(\Gamma C_A+C_B\right)-\Gamma-1\right]^2+2M_0\Gamma}. \label{psi}
\end{equation}
In Eqs. (\ref{LAALABLBB}) and (\ref{psi}), $\Gamma=\Gamma_A/\Gamma_B$, $\Gamma_A$ and $\Gamma_B$ are the jump frequencies of species A and B, respectively,
$M_0=2f_0/\left(1-f_0\right)$, $f_0$ and $\lambda$ are the tracer correlation factor and the 
dimensionless geometric factor for fcc lattice, respectively, and $\bm{W}$ the diagonal, transversely isotropic tensor,
\begin{equation}
\bm{W} = 
\begin{pmatrix}
1 & 0\\
0 & \Lambda
\end{pmatrix},
\label{DiffTensor}
\end{equation}
where the dimensionless $\Lambda$ measures the difference in the diffusion strength along the $x$ and $y$ axes. Specifically, if we let $D_x$ and $D_y$ 
be the diffusivities along the $x$ and $y$ axes, respectively, then $D_y=\Lambda D_x$.   
Thus the multiplication of the kinetic transport coefficients by $\bm{W}$ allows to account for the possible directional 
diffusional anisotropy in the xy-plane. 
When $\Lambda=1$, $\bm{W}=\bm{I}$,
where $\bm{I}$ is the identity tensor, and the diffusion becomes isotropic. 

Equations (\ref{sumC}), (\ref{divV})-(\ref{j-eq2}), (\ref{base_eq4})-(\ref{DiffTensor}) constitute our dimensional model, with $\nu_A=\nu_B=\nu$ in Eq. (\ref{base_eq4}).
To render this model dimensionless, 
we chose $a$ as the length scale
and $\bar t=k T\nu/\left(\gamma_B \lambda \Gamma_B \right)$ as the time scale. After elimination of $C_A$ using Eq. (\ref{sumC}), we arrive to the final 
dimensionless model for the spatio-temporal dynamics of the concentration of B atoms:
\begin{equation}
\frac{\partial C_B}{\partial t} =   -\bm{\nabla} \cdot \left[F_e \bm{\Omega}_B-\bm{\Sigma}_B+
\left(\bm{\Sigma_A}+\bm{\Sigma_B}
-F_e \bm{\Omega}_A -F_e \bm{\Omega}_B\right)C_B\right],  \label{C-eq32}
\end{equation}
\begin{eqnarray}
\bm{\Sigma}_A &=&\bar L_{AA} \bm{\nabla}_\Lambda \mu_A + \bar L_{AB} \bm{\nabla}_\Lambda \mu_B,\quad 
\bm{\Sigma}_B = \bar L_{AB} \bm{\nabla}_\Lambda \mu_A + \bar L_{BB} \bm{\nabla}_\Lambda \mu_B, \label{SigmaAB} \\ \nonumber \\
\bm{\Omega}_A &=& \left[Q\bar L_{AA}+\bar L_{AB}\right]\begin{pmatrix}
\cos{\phi_E}\\
\Lambda \sin{\phi_E}
\end{pmatrix},\quad \bm{\Omega}_B = \left[Q\bar L_{AB}+\bar L_{BB}\right]\begin{pmatrix}
\cos{\phi_E}\\
\Lambda \sin{\phi_E}
\end{pmatrix}, \label{OmegaAB}
\end{eqnarray} 
\begin{equation}
\bar L_{AA}=\Gamma \left(1-C_B\right) \left(1-\frac{2\Gamma C_B}{\psi}\right),\quad 
\bar L_{AB}=\frac{2}{\psi}\Gamma C_B \left(1-C_B\right),\quad
\bar L_{BB}=C_B \left(1-\frac{2 \left(1-C_B\right)}{\psi}\right), \label{Ls}
\end{equation}
\begin{equation}
\mu_A = -C_B\frac{\partial \gamma}{\partial C_B}+\xi \bm{\nabla}^2 C_B,\quad
\mu_B = \left(1-C_B\right)\frac{\partial \gamma}{\partial C_B}-\xi \bm{\nabla}^2 C_B, 
\label{base_eq41}
\end{equation}
\begin{equation}
\gamma = G \left(1-C_B\right) + C_B  + N\left[\left(1-C_B\right)\ln \left(1-C_B\right)+C_B\ln C_B+H C_B \left(1-C_B\right)\right],
\label{nondim_gamma}
\end{equation}
\begin{eqnarray}
\psi &=& \frac{1}{2}\left(M_0+2\right)\left(\Gamma \left(1-C_B\right)+C_B\right)-\Gamma-1+2\left(\left(1-C_B\right)+\Gamma C_B\right)+ \nonumber \\
&&\sqrt{\left[\frac{1}{2}\left(M_0+2\right)\left(\Gamma \left(1-C_B\right)+C_B\right)-\Gamma-1\right]^2+2M_0\Gamma}, \label{psi_nondim}
\end{eqnarray}
\begin{equation}
\bm{\nabla}_\Lambda=
\begin{pmatrix}
\partial_x\\
\Lambda\partial_y
\end{pmatrix}.
\label{misc} 
\end{equation}
Here, $\bm{\nabla}_\Lambda$ is the anisotropic gradient operator, which reduces to the $\bm{\nabla}$ operator when diffusion
is isotropic. The fluxes $\bm{\Sigma}_A$ and $\bm{\Sigma}_B$ are due to the gradients of the chemical potentials, whereas 
the fluxes $\bm{\Omega}_A$ and $\bm{\Omega}_B$ are the electromigration fluxes. Since the anisotropy matrix $\bm{W}$ enters the three kinetic transport
coefficients in Eqs. (\ref{j-eq2}), all fluxes are anisotropic either due to presence of $\bm{\nabla}_\Lambda$,
or due to presence of the column vector $\begin{pmatrix}
\cos{\phi_E}\\
\Lambda \sin{\phi_E}
\end{pmatrix}
$. 
Also, $\bar L_{AA},\ \bar L_{AB}$ and $\bar L_{BB}$ are the dimensionless, and now isotropic, kinetic transport coefficients.
\footnote{A mean drift velocity term $-F_e \left(\bm{\Omega}_A +\bm{\Omega}_B\right)\cdot \bm{\nabla}C_B$ is added to the right-hand side of Eq. (\ref{C-eq32})
prior to computation \cite{OPLE}.}

Eqs. (\ref{C-eq32})-(\ref{misc}) contain ten dimensionless parameters. Parameters $\Gamma,\ \Lambda,\ H,\ M_0,\ \phi_E$ were introduced above. 
Note, when $H>2$, the graph of $\gamma$ is a double-well curve, 
which typically results in the spinodal decomposition of the alloy. 
The condition $H>2$ is equivalent to $T<T_c=\alpha_{int}/2k\nu$ \cite{LuKim}, and for the parameters in Table \ref{T1}, $T_c=455$K. 
Other parameters are:
$G=\gamma_A/\gamma_B$, the ratio of the surface energies of the components; $N=k T\nu /\gamma_B$: the alloy entropy; $\xi=\epsilon /a \gamma_B$: the
Cahn-Hilliard gradient energy coefficient; $Q=q_A/q_B> < 0$: the ratio of the effective charges; $F_e=q_B a \nu V/ L \gamma_B > <0$: 
the electric field strength.

\begin{table}[!ht]
\centering
{\scriptsize 
\begin{tabular}
{|c|c|c|c|}

\hline
				 
			\rule[-2mm]{0mm}{6mm} \textbf{Physical parameter}	 & \textbf{Typical value} & \textbf{Dimensionless parameter}	 & \textbf{Typical value}  \\
			\hline
                        \hline
			\rule[-2mm]{0mm}{6mm} $L$ & $10^{-5}$ cm (100 nm) \cite{GSASF} & $\Gamma$ & 2\\
			\hline
                        \rule[-2mm]{0mm}{6mm} $a$ & $3.3\times 10^{-8}$ cm \cite{CRR}  (0.33 nm) & $\Lambda$ & 1 \\
			\hline
				\rule[-2mm]{0mm}{6mm} $\nu=a^{-2}$ & $9.18\times 10^{14}$ cm$^{-2}$ & $H$  & 2.27  \\
			\hline
            \rule[-2mm]{0mm}{6mm} $\lambda$ & $1/6$ \cite{Manning1} & $M_0$ & 7.153\\
                        \hline
		    \rule[-2mm]{0mm}{6mm} $f_0$ & $0.7815$  \cite{Manning1} & $\phi_E$ & $\pi$\\
                        \hline
			\rule[-2mm]{0mm}{6mm} $T$ & 400 K \cite{AAFS} & $G$ & 1.966\\
			\hline
			    \rule[-2mm]{0mm}{6mm} $q_A,\ q_B$ & $20|e|=10^{-8}$  statC \cite{BGWZ,REW} & $N$ &  0.044\\
			\hline
			    \rule[-2mm]{0mm}{6mm} $V$ & $2.25\times 10^{-2}$V   & $\xi$ &  0.31\\	
			\hline
			\rule[-2mm]{0mm}{6mm} $\alpha_{int}$ & $117$  erg$/$cm$^2$  & $Q$ & 1 \\ 
			\hline
			    \rule[-2mm]{0mm}{6mm} $\gamma_A$ & $2.3\times 10^3$  erg$/$cm$^2$ \cite{VRSK}  & $F_e$ & $0.583$ \\ 
				\hline
				\rule[-2mm]{0mm}{6mm} $\gamma_B$ & $1.17\times 10^3$  erg$/$cm$^2$ \cite{VRSK}  & & \\
				\hline
			    \rule[-2mm]{0mm}{6mm} $\epsilon$ & $1.2\times 10^{-5}$ erg$/$cm  \cite{Hoyt} & & \\
			\hline

\end{tabular}}
\caption[\quad Physical parameters]{Physical and dimensionless parameters for fcc Ag/Pt(111) alloy. 
$a$ is the computed covalent diameter of the Ag atom, which
corresponds to a B atom in the model. For the discussion of $\alpha_{int}$ see Sec. \ref{Disc}. 
}
\label{T1}
\end{table}

\begin{figure}[H]
\vspace{-0.2cm}
\centering
\includegraphics[width=3.5in]{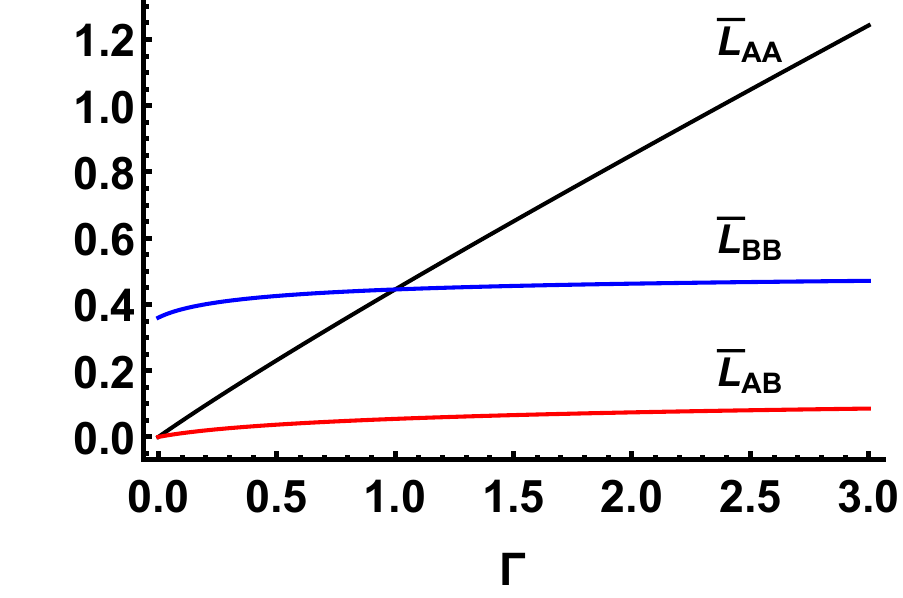}\hspace{0.5cm}
\vspace{-0.15cm}
\caption{Kinetic transport coefficients $\bar L_{AA},\ \bar L_{AB}$ and $\bar L_{BB}$ at $C_B=0.5$. At $\Gamma=2$ (A is fast diffuser), $\bar L_{AA}=0.851$, $\bar L_{BB}=0.463$,
$\bar L_{AB}=0.074$. At $\Gamma=0.2$ (B is fast diffuser), $\bar L_{AA}=0.096$, $\bar L_{BB}=0.4$,
$\bar L_{AB}=0.02$. These values mildly change as $C_B$ is varied, and they are provided only for comparison of the magnitudes, not for computation.
}
\label{Fig2}
\end{figure}

The dimensionless kinetic transport coefficients are plotted in Fig. \ref{Fig2}. 
Note that at $\Gamma\rightarrow 0$, meaning $\Gamma_B \gg \Gamma_A$, 
only $\bar L_{BB}$ is not zero, thus A atoms are immobile. And at $\Gamma\rightarrow \infty$, meaning $\Gamma_A \gg \Gamma_B$, 
$\bar L_{BB}$ and $\bar L_{AB}$ are negligible in comparison to $\bar L_{AA}$, thus B atoms are effectively immobile. 
At $\Gamma=1$ $(\Gamma_A =\Gamma_B)$ , $\bar L_{AA}=\bar L_{BB}$. 

\begin{figure}[H]
\vspace{-0.2cm}
\centering
\includegraphics[width=3.5in]{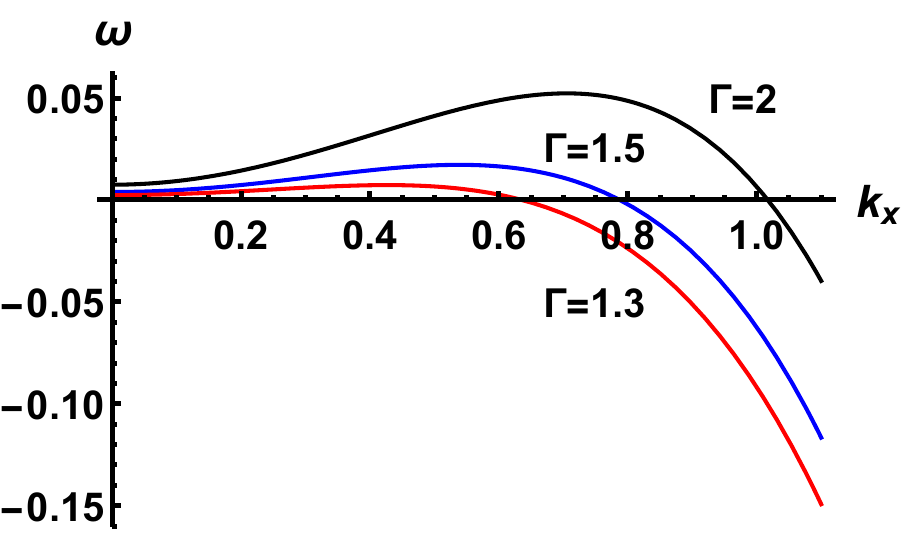}\hspace{0.5cm}
\vspace{-0.15cm}
\caption{The perturbation growth rate from LSA of Eq. (\ref{C-eq32}) at $F_e=0$.
}
\label{Fig2_1}
\end{figure}

Fig. \ref{Fig2_1} shows the graphs of the perturbation growth rate $\omega$ from the linear stability analysis (LSA) of Eq. (\ref{C-eq32}) 
(linearized about $C_B=0.5$). As calculated, $\omega$ is a function of the perturbation wavenumbers $k_x$ and $k_y$ in the $x$ and $y$ directions, respectively; 
here we plotted the cross-section, $\omega(k_x,0.2)$ for the parameters in Table \ref{T1}, except $V=F_e=0$ (no electromigration). 
Here, due to the symmetry in the absence of anisotropy ($\Lambda=1$), a graph of $\omega(c,k_y),\; c=const.$ is the same as a graph of $\omega(k_x,c)$. 
($\omega$ is complex-valued in the presence of electromigration, see for example Ref. \cite{MySurfSci}, but the real part of $\omega$ is not affected by 
electromigration.) The maximum of the plotted $\omega$ curve shifts toward larger wavelengths as $\Gamma$ decreases. At $\Gamma\le 0.97$ the entire curve 
is located below the $k_x$ axis for any fixed $k_y$ ($\omega<0$), thus for such $\Gamma$ values any small perturbation of a homogeneous state A$_{0.5}$B$_{0.5}$ would decay in time, and 
the homogeneous state would be restored. In other words, phase separation is absent and there is phase stability. Note that 
this situation occurs at values of the thermodynamic parameters $H,\; G$ and $N$ that in a Cahn-Hilliard-type model with a generic diffusion account 
would result in the spinodal instability of the A$_{0.5}$B$_{0.5}$ alloy; 
thus at $\Gamma\le 0.97$ the vacancy-mediated diffusion eliminates the said instability. For this reason, in the next section only $\Gamma > 0.97$ is considered. 
In Section \ref{Disc} we provide the insight into the origin of this interesting and important effect.
\section{Computational results}
\label{CompositionEvolve}

\begin{figure}[H]
\vspace{-0.2cm}
\centering
\includegraphics[width=6.5in]{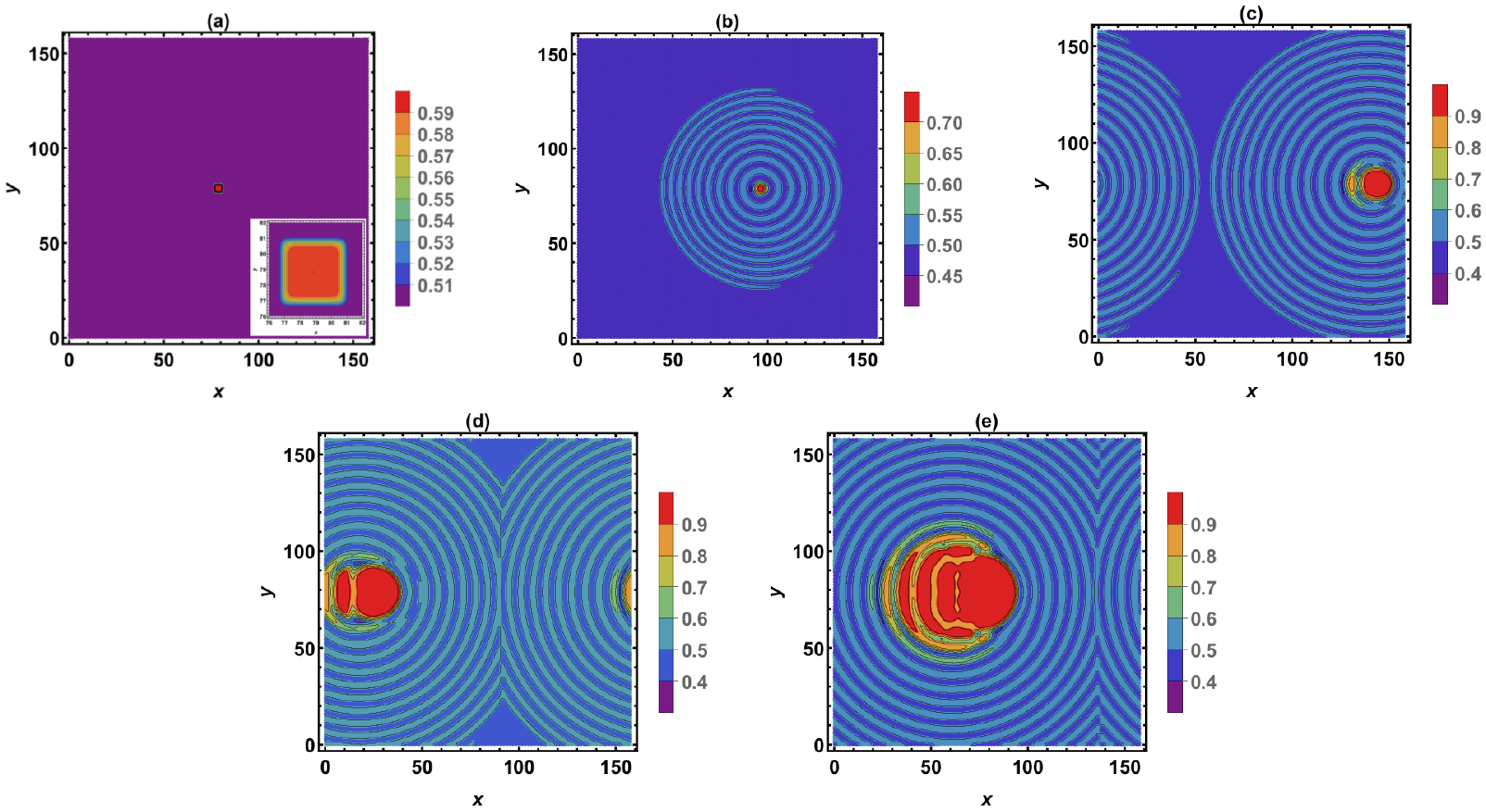}
\vspace{-0.15cm}
\caption{$C_B$ at (a): $t=0$ (the initial condition), (b): $t=20$, (c): $t=70$, (d): $t=110$, 
(e): $t=160$.  The initial cluster is enlarged in the inset of (a). Values of all parameters are in Table \ref{T1}.
}
\label{Fig3}
\end{figure}

Computations of Eq. (\ref{C-eq32}) are performed on the fixed, square, bi-periodic domain of the size $158a$, i.e. 158 lattice spacings, 
which approximately equals to one-half the width of the terrace ($L/2a=152$). For the parameters in Table \ref{T1}, 158 lattice spacings 
equals to $13\lambda_{max}$, where $\lambda_{max}$ is the most dangerous wavelength from LSA.
The initial condition $C_B(x,y,0)$ is chosen in the form of a local composition inhomogeneity
positioned at the square center; see Fig. \ref{Fig3}(a). 
Within this initial cluster (size: $4\times 4$ lattice spacings), the concentration of B atoms is increased to 60\% (see the color bar). This is 10\% higher than outside of the cluster, where
the species A and B have equal concentrations, that is, 50\%A/50\%B.

Panels (b)-(e) of Fig. \ref{Fig3} show the transport (drift) of the cluster due to the electromigration effect caused by the electric field $\bm{E}$ 
in the $180^\circ$ direction (see Fig. \ref{Fig1}). The dimensionless parameters are taken from Table \ref{T1}. After drifting across the half of the computational domain,
the cluster re-emerges at the left boundary of the domain due to the periodic boundary condition. By $t=160$ the center of the cluster covered the 
distance roughly equal to the domain width of 158 lattice spacings.  
To obtain a dimensional cluster's speed, first we estimate value of $\Gamma_B$ from the Arrhenius-type expression $\Gamma_B=\sigma \exp{\left( -E_a/kT \right)}$, 
where $\sigma=10^{13}$Hz is the attempt frequency and $E_a\simeq 1$eV the activation  energy \cite{BEFBK}.\footnote{Since we could not find the activation 
energy of vacancy diffusion for AgPt(111) surface alloy, the cited value is for vacancy diffusion on a clean Cu surface.}
This gives $\Gamma_B=4$Hz, and then at the parameters values in Table \ref{T1} the time scale $\bar t=0.0675$s. Using this value, $t=158$ corresponds to 11 s, 
and the cluster's speed is 14 lattice spacings/s.  As the cluster drifts, due to developing phase separation instability
the B (Ag) atoms from the surrounding surface layer "flow" into the cluster and correspondingly the A (Pt) atoms flow out. Thus the cluster grows with time. 
In Fig. \ref{Fig3}(e), the dimensions of the cluster's core, e.g. the area whose shape resembles a semi-circle, are 29$\times$36 lattice spacings.  Close to the end of the run, the composition is 
100\% B phase, see the color bars in Fig. \ref{Fig3}(d,e). 
The development of the instability, which takes place independently of, and alongside with the drift, 
also results in the repeated formation and breaking of the "neck" at the trailing edge of the cluster. Small clusters that are shedded this way from the 
cluster's core evolve into B-rich arcs that drift behind the core. Drawing the analogy from biology, the shape of a medium-sized cluster's core, neck, and a small 
cluster at the end of a neck (Fig. \ref{Fig3}(d)) 
resembles a jellyfish, with the key distinction that the latter actively self-propells in water, whereas the cluster passively drifts in the electron wind.

%
\begin{figure}[H]
\vspace{-0.2cm}
\centering
\includegraphics[width=4.5in]{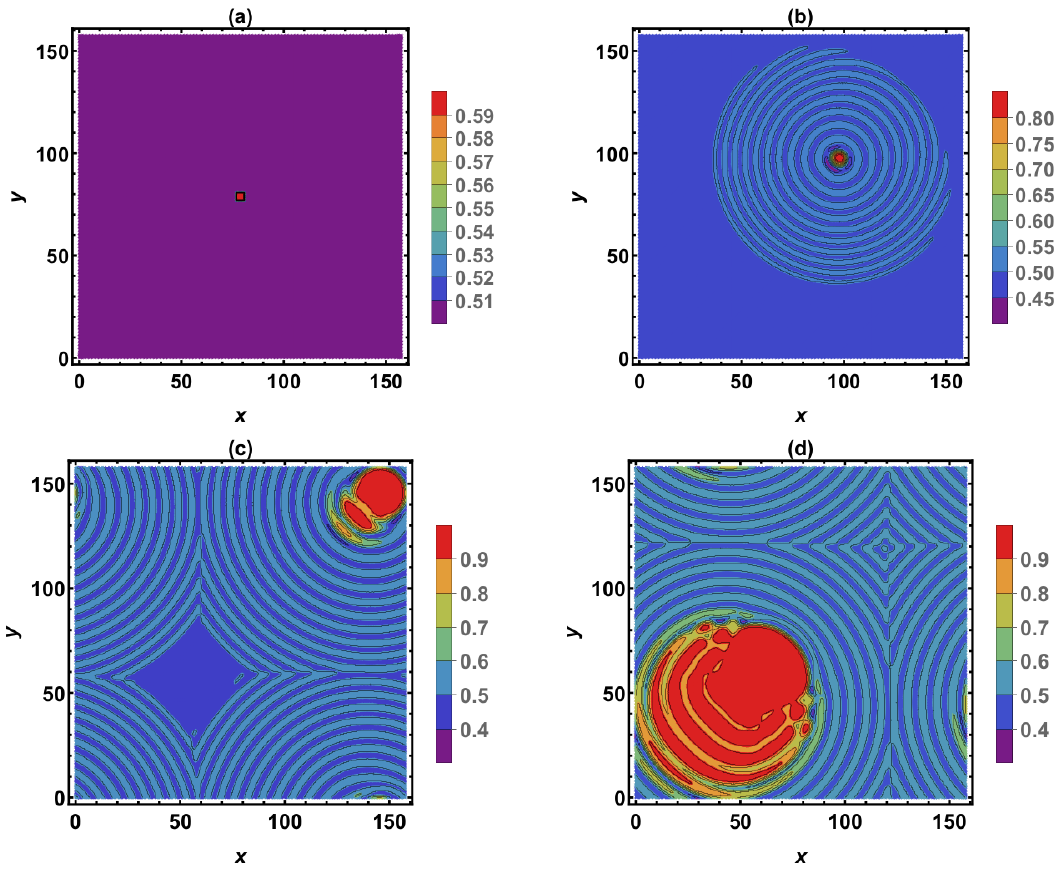}
\vspace{-0.15cm}
\caption{$C_B$ at (a): $t=0$ (the initial condition), (b): $t=30$, (c): $t=100$, (d): $t=200$. 
$\phi_E=5\pi/4$, other parameters as in Table \ref{T1}.
}
\label{Fig4}
\end{figure}

In Fig. \ref{Fig4} the direction of the electric field, and of the current, is $225^\circ$, and all other parameters are still as in Table \ref{T1}.
The change in the electic field
direction results in the cluster drifting along the square's diagonal with the same speed and other features seen in Fig. \ref{Fig3}. From Figures \ref{Fig3}
and \ref{Fig4} it is clear that the electromigration very reliably transports 
the cluster carrying the effective positive charge, across the terrace in the direction that is the opposite to the direction of the current. 

In the remaining computations (except the results shown in Figures \ref{Fig7} and \ref{Fig8}) we fix the electric field direction to $180^\circ$, as in Fig. \ref{Fig3}, and study the effects of other parameters.

\begin{figure}[H]
\vspace{-0.2cm}
\centering
\includegraphics[width=5.5in]{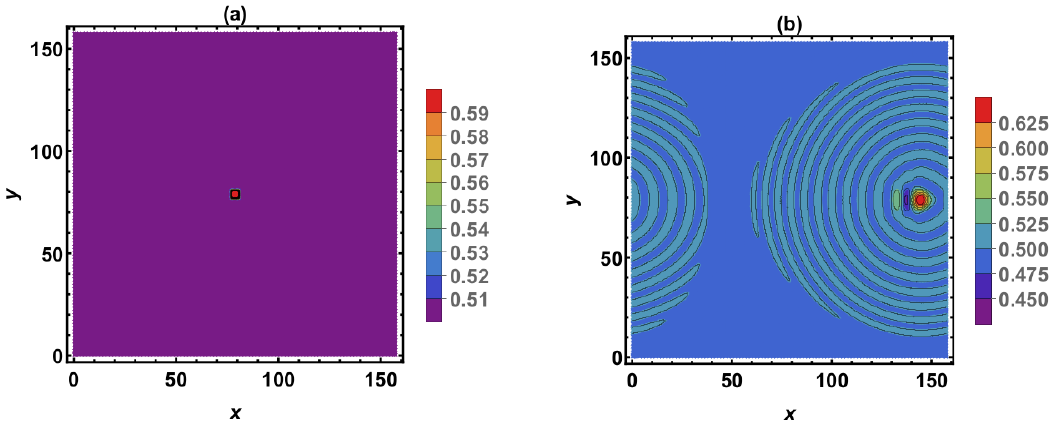}
\vspace{-0.15cm}
\caption{$C_B$ at (a): $t=0$ (the initial condition), (b): $t=100$.  $\Gamma=1.3$, other parameters as in Table \ref{T1}.
}
\label{Fig5}
\end{figure}

Next, Fig. \ref{Fig5} shows the effect of varying the parameter $\Gamma$. As $\Gamma$ is decreased from 2 to 1.3, the speed of the 
cluster decreased by over 40\%, compare to Fig. \ref{Fig3}(c). Since $\Gamma_B$ is kept constant (notice that $\Gamma_B$ enters
only the time scale, which has not changed), the stated decrease of $\Gamma$ is due to the decrease of $\Gamma_A$. Thus the slow-down of the drift
is attributed to the smaller hop frequency of A atoms, which resulted in the decrease of the kinetic transport coefficients, see Fig. \ref{Fig2} 
and its caption. Despite the slower cluster drift its enrichment by B atoms is less efficient than at $\Gamma=2$, and the cluster's size and 
concentration are smaller than in Fig. \ref{Fig3}(c). This is consistent with the LSA, see the discussion of Fig. \ref{Fig2_1} 
(the phase separation is partially or completely mitigated by decreasing $\Gamma$). Computation on this fixed domain at $\Gamma<1.3$ still resulted in
the cluster drift, but with the phase separation diminishing with time, until the cluster dissolved into the surface.   

\begin{figure}[H]
\vspace{-0.2cm}
\centering
\includegraphics[width=7.0in]{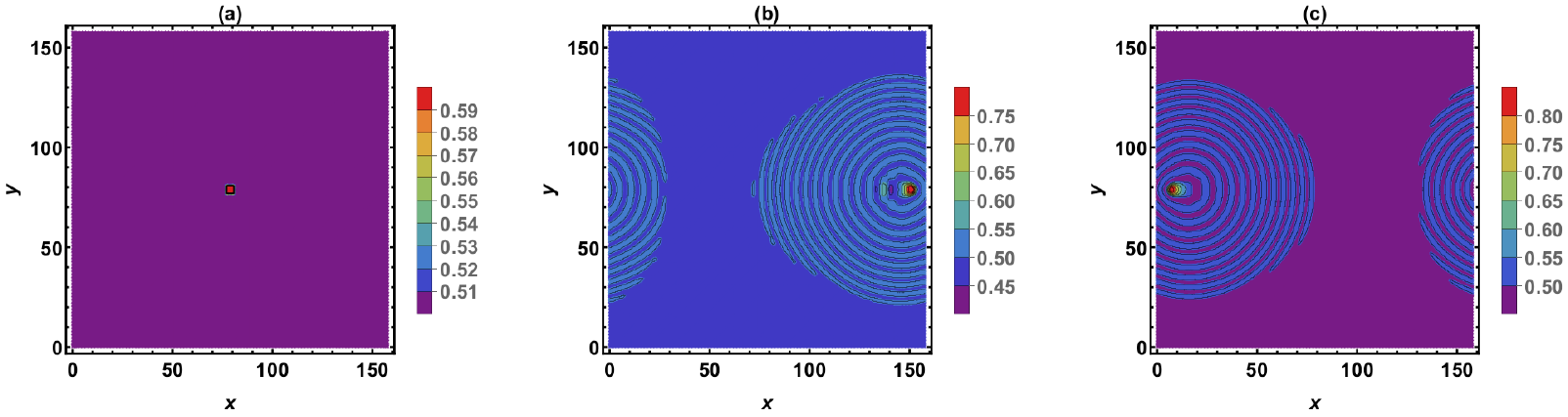}
\vspace{-0.15cm}
\caption{$C_B$ at (a): $t=0$ (the initial condition), (b): $t=70$, $q_A=5q_B>0$ ($Q=5$), other parameters as in Table \ref{T1}, 
(c): $t=80$, $q_A=-5q_B<0$ ($Q=-5$), other parameters as in Table \ref{T1}.
}
\label{Fig6}
\end{figure}

In Fig. \ref{Fig6}(b) it is seen that larger (but still positive) effective charge of A atoms $\left(q_A\rightarrow 5q_A\right)$ 
results in smaller cluster and smaller $C_B$ within the cluster, while the drift speed is not affected 
(compare to Fig. \ref{Fig3}(c)). Less efficient phase separation here is the purely \emph{kinetic} effect of electromigration, since a thermodynamic force
that causes phase separation remains constant. To understand the origin of this kinetic effect, observe in Eq. (\ref{OmegaAB}) that the parameter $Q$ 
enters asymmetrically the expressions for electromigration fluxes $\bm{\Omega}_A$ and $\bm{\Omega}_B$. As seen in Fig. \ref{Fig6}(c) and as expected, 
changing not only the magnitude, but also the sign of $q_A$ results in the reversal of the drift direction, with other features of Fig. \ref{Fig6}(b) 
remain essentially unchanged. 

\begin{table}[!ht]
\centering
{\scriptsize 
\begin{tabular}
{|c|c|c|c|}

\hline
				 
			\rule[-2mm]{0mm}{6mm} \textbf{Applied voltage (V)}	 & \textbf{Dimensionless speed} & \textbf{Dimensionless size}	 & \textbf{Composition (\% B})  \\
			\hline
                        \hline
			\rule[-2mm]{0mm}{6mm} 0.005 & 0.33 & $38\times 33$ & 95\\
			\hline
                        \rule[-2mm]{0mm}{6mm} 0.01 & 0.47 & $26\times 29$ & 95\\
			\hline
				\rule[-2mm]{0mm}{6mm} 0.015 & 0.63 & $19\times 23$ & 95\\
			\hline
				\rule[-2mm]{0mm}{6mm} 0.02 & 0.825 & $15\times 15$ & 95\\
                        \hline
            \rule[-2mm]{0mm}{6mm} 0.0225 & 0.94 & $12\times 11$ & 95\\
                        \hline
		    \rule[-2mm]{0mm}{6mm} 0.025 & 1.01 & $14\times 13$ & 95\\
                        \hline

\end{tabular}}
\caption[\quad Cluster stats]{Cluster speed, size, and composition. The cluster's core size and composition are stated at a time when the cluster 
reaches the right boundary of the computational domain. Note that values of the applied voltage can be converted into values of the current density 
using the calculation shown in Sec. \ref{Disc}.
}
\label{T2}
\end{table}  

Again for the parameters in Table \ref{T1}, Table \ref{T2} shows that the cluster speed increases linearly with the increase of the applied voltage, 
whereas the cluster size decreases. 
(The higher the voltage and thus the speed, the less time is afforded to the phase separation instability 
to replace A atoms on the cluster boundary with B atoms.) 
The linear fit to speed vs. voltage data is $v=34 V +0.16$; the linear law breaks down at small $V$. We found that at $V=0.0005$V the cluster is essentially immobile on the time scale $0\le t\le 500$.
That is, it stays very close to the center of the computational domain and grows until it fills the entire domain area. 

%
\begin{figure}[H]
\vspace{-0.2cm}
\centering
\includegraphics[width=7.0in]{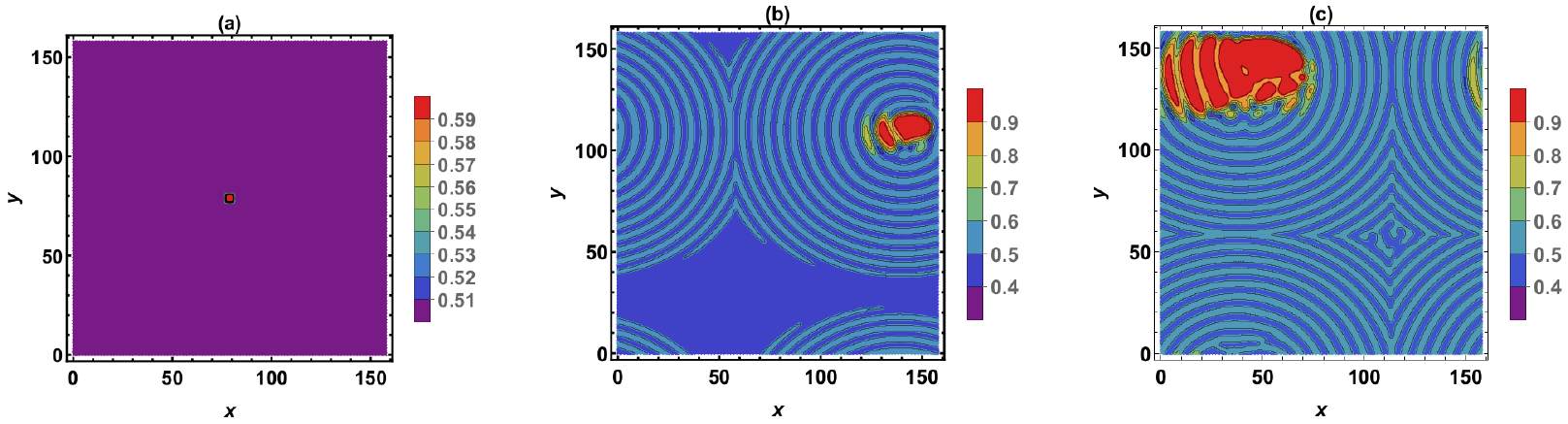}
\vspace{-0.15cm}
\caption{$C_B$ at (a): $t=0$ (the initial condition), (b): $t=10$,  (c): $t=190$. $\Lambda=0.5$, $\phi_E=5\pi/4$, other parameters as in Table \ref{T1}. 
}
\label{Fig7}
\end{figure}
\begin{figure}[H]
\vspace{-0.2cm}
\centering
\includegraphics[width=7.0in]{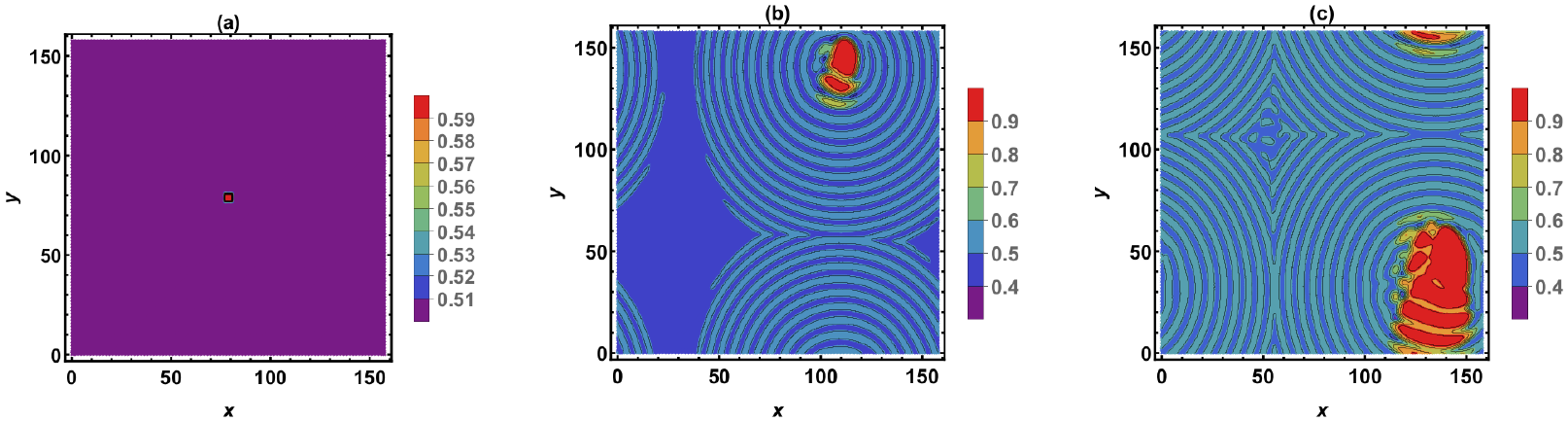}
\vspace{-0.15cm}
\caption{$C_B$ at (a): $t=0$ (the initial condition), (b): $t=50$,  (c): $t=90$. $\Lambda=2$, $\phi_E=5\pi/4$, other parameters as in Table \ref{T1}. 
}
\label{Fig8}
\end{figure}

Lastly, in Figures \ref{Fig7}, \ref{Fig8} we clarify the impact of the diffusion anisotropy. The computation is done with $\Lambda=0.5$ and $\Lambda=2$,
respectively, using $\phi_E=225^\circ$. The results should be compared to Fig. \ref{Fig4}.  
It is clear that strong anisotropy has the detrimental effect on the target direction of the drift.\footnote{Ref. \cite{GSBH} provides the ratio close to 0.25 of 
the number of the diagonal hops over the distance $\sqrt{2}a$ to the number of hops over the distance $a$ along the $x$ or $y$ direction. This seems to indicate that 
diffusion anisotropy is typically manifested in vacancy-mediated surface diffusion.} At $\Lambda=0.5$ the diffusion
in the $x$-direction is twice as strong as in the $y$-direction, which results in the cluster straying, by about 30$^\circ$ clockwise, from the path along 
the square's diagonal. At $\Lambda=2$, when the diffusion
in the $y$-direction is twice as strong as in the $x$-direction,  the deviation from the path along the diagonal is roughly 30$^\circ$ counter-clockwise. 
The shape of the cluster is also impacted: losing a near-circular symmetry, the cluster's core is elongated and bended.

\section{Discussion}
\label{Disc}

We begin this section with estimates of the electromigration current density, the temperature rise of the film, and the enthalpy of mixing 
of the surface alloy.

Let $\rho$ be the resistivity of Ag$_x$Pt$_{1-x}$ film. Then for the electromigration current density $j$ we have:
\begin{equation}
j=E_0/\rho=V/(L\rho),\quad \rho=(1-x)\rho_{Pt}+x \rho_{Ag},
\end{equation}
where $\rho_{Ag}=2.24\times 10^{-6}\Omega$cm,  $\rho_{Pt}=14.7\times 10^{-6}\Omega$cm at 400K \cite{D1,D2}. Substituting these values, $x=0.5$, and $V$, $L$ 
from Table \ref{T1}, gives $\rho=8.47\times 10^{-6}\Omega$cm, $j=2.65\times 10^8$ V$/\Omega$cm$^2$. 

The temperature rise of a thin metal film on a thick dielectric substrate due to electromigration currrent can be calculated using Black's formula \cite{Black}:
\begin{equation}
\Delta T = \rho j^2 h_{mf} h_{ds}/\chi_{ds},
\end{equation}
where $h_{mf}$ and $h_{ds}$ are the thickness of a film and of a substrate, respectively, and $\chi_{ds}$ is the thermal conductivity of a substrate.
Using $h_{mf}=20$nm, $h_{ds}=200$nm, $\rho$, $j$ as calculated above, and $\chi_{ds}=\chi_{Pt}\approx 75$ W$/$m K at 400K \cite{D3,D4} gives $\Delta T = 30$K.
\footnote{Thermal conductivity of bulk dielectric materials varies greatly, from approximately 1 W$/$m K for a-SiO$_2$ 
to approximately 2000 W$/$m K at 400K for diamond.}
Thus despite the temperature rise from 400K to 430K due to electromigration, the film is still below the critical temperature $T_c\sim 455$K of the spinodal decomposition, 
provided that the chosen value of the enthalpy of mixing $\alpha_{int}=0.1\gamma_B$ in Table \ref{T1} is order-of-magnitude accurate. 

To confirm this,
we convert $\alpha_{int}=27\times 10^{-3}$ eV/atom for Ag$_{0.5}$Pt$_{0.5}$ alloy \cite{SCP} to a value in units of erg$/$cm$^2$. 
Using $a/2$ from Table \ref{T1} for the radius of Ag atom, and the radius of Pt atom $1.77\times 10^{-8}$ cm \cite{CRR}, we obtain the average radius  
$r_{av}=1.71\times 10^{-8}$cm. Asssuming first that the packing of the surface layer is by solid discs of the radius $r_{av}$, representing the surface atoms,
we obtain $1.1\times 10^{15}$ for the number of atoms in one square centimeter of the surface, and then  $\alpha_{int}=47$ erg$/$cm$^2$. Alternatively,
assuming that the packing of the surface layer is by solid squares of the side $r_{av}$, we obtain $3.4\times 10^{15}$ for the number of atoms in one square centimeter 
of the surface, and then  $\alpha_{int}=148$ erg$/$cm$^2$. The chosen value $\alpha_{int}=117$ erg$/$cm$^2$ in Table \ref{T1} falls neatly between these 
bounds.

For completeness, in the following remark we compare data from literature on diffusion rates of various atom exchange processes.

For example, according to first principle calculations by Grant \textit{et al.} \cite{GSBH} for
PdCu(001) surface alloy, 
the activation energy for Pd-Cu adatom exchange is 0.95 eV, while the activation energy for vacancy diffusion is only 0.466 eV. Here, adatom exchange  
means that Pd atom pops out from the surface layer onto the surface, diffuses on the surface, and then reincorporates back into the surface layer.
Thus based on these values vacancy-mediated diffusion is expected to predominate. As another example, Anderson \textit{et al.} \cite{ABFSK} found that 
Pb atoms embedded in the Cu(111) surface suppress surface diffusion rates by several orders of magnitude - the surface diffusion 
barrier for Cu adatom diffusion increases from 0.8 eV to 0.9 eV with 0.11 ML Pb, and to 1.2 eV with 0.22 ML Pb. This is another sign that 
vacancy-mediated diffusion is predominant.  

To end this section, next we provide the formal explanation of the suppression of the phase separation at the small values of the ratio of the hop frequencies
$\Gamma$, as seen in LSA (see section \ref{Model}).

It is sufficient for this purpose to consider the simplified Eq. (\ref{C-eq32}),
\begin{equation}
\frac{\partial C_B}{\partial t} =   \left(1-C_B\right)\bm{\nabla}\cdot \left(\bar L_{AA} \bm{\nabla} \mu_A\right) - 
C_B\bm{\nabla}\cdot \left(\bar L_{BB} \bm{\nabla} \mu_B\right).
\label{C-eq32simple}
\end{equation}
Here, the electric field parameter $F_e=0$, the convective terms are omitted, isotropic diffusion is assumed, and the contributions proportional 
to $\bar L_{AB}$ also are omitted (since it is seen in Fig. \ref{Fig2} that $\bar L_{AB}\ll \bar L_{AA}, \bar L_{BB}$ for any $\Gamma$). Also, we can
``freeze" the coefficients of two terms on the right-hand side of  Eq. (\ref{C-eq32simple}) by fixing $C_B$; we need to take $C_B=1/2$, since this value is used in LSA
and computation for the initial composition. Via the chemical potentials $\mu_{A,B}$ (Eqs. (\ref{base_eq41})), Eq. (\ref{C-eq32simple}) contains only 
the thermodynamic effects, which are affected by vacancy diffusion kinetics via the composition-dependent coefficients $\bar L_{AA}$ and $\bar L_{BB}$.
We now simplify the equation further by assuming that $\bar L_{AA}$, $\bar L_{BB}$ are constant; and take $\bar L_{AA}=\theta \bar L_{BB}$, $0<\theta<1$.
This choice of the interval for $\theta$ corresponds to the relation between $\bar L_{AA}$ and $\bar L_{BB}$ in Fig. \ref{Fig2} at $\Gamma<1$. 
Next, taking $\xi=0$ without loss of generality, and substituting $\mu_{A,B}$ gives
\begin{equation}
\frac{\partial C_B}{\partial t} =   \frac{\bar L_{BB}}{2}\bm{\nabla}^2\left[C_B(1-\theta)-1\right]\frac{\partial \gamma}{\partial C_B}.
\label{C-eq32simple1}
\end{equation}
We again fix $C_B$ to a constant in the interval [0,1] (a particular value 1/2 may be chosen), then it is obvious that $C_B(1-\theta)-1= -w$, where 
$w>0$ is another constant. Finally, Eq. (\ref{C-eq32simple1}) reads
\begin{equation}
\frac{\partial C_B}{\partial t} =   \frac{-w\bar L_{BB}}{2}\bm{\nabla}^2\frac{\partial \gamma}{\partial C_B},
\label{C-eq32simple2}
\end{equation}
where $\gamma$ is given by Eq. (\ref{nondim_gamma}). The LSA of Eq. (\ref{C-eq32simple2}) about $C_B=1/2$ gives $\omega=N w \bar L_{BB}(2-H)(k_x^2+k_y^2)$. 
Thus $\omega<0$ for all wavenumbers at $H>2$ (e.g., all small perturbations of the homogeneous state $C_B=1/2$ decay). But $H>2$ is the condition of 
the spinodal instability. This calculation makes it clear that the relation
$\bar L_{AA} < \bar L_{BB}$, which results at $\Gamma <1$, and the particular choice of the alloy composition A$_{0.5}$B$_{0.5}$, give rise to the
the suppression of the phase separation.


\section{Summary}
\label{Summ}

To summarize, we suggested harnessing a phase separation instability and surface electromigration to enable transport across terraces of A or B 
atom clusters embedded in the topmost surface layer of a thermodynamically unstable surface binary alloy (e.g., the alloy's enthalpy of mixing is positive). 
Our synthetic continuum model rests on the prior assessement from STM experiments and ab initio computations that vacancy-mediated diffusion is 
often predominant for surface alloys. Computations are executed with parameters that closely mimick AgPt(111) surface alloy, which is known to be 
thermodynamically unstable and form the embedded Ag clusters.
The model is developed by first reducing the classical model of a vacancy-mediated diffusion in the bulk 3D binary 
alloy couples to an alloyed, dense 2D surface layer, and then combining with the consistent, irreversible thermodynamics-based model of electromigration 
in alloys. The model allows to study the effects that are often overlooked, such as the diffusion anisotropy and the unequal and 
sign-wise different effective charges of A and B atoms. (To our knowledge, there is no published assessement of the latter effect in 
any context even for bulk metallic alloys.) For this application, we find that the unequal effective charges may kinetically slow down the phase separation 
in the binary surface layer. Phase separation also is slowed down or brought to a standstill when the ratio of the hop frequencies of A and B atoms
is in certain range. Increasing the applied voltage results in a smaller cluster size and the linear increase of its speed, whereas the purity is not significantly
affected. Overall, the model's computations demonstrate the feasibility of accomplishing the stated goal, e.g. enabling the transport across 
terraces of the embedded A or B atom clusters. Of course, to be successful, a material system chosen for the experimental validation will require 
carefull optimization and calibration across a multi-parameter landscape.




\end{document}